\documentclass{article}
\usepackage[accepted]{vietnam} 
\usepackage{natbib}
\usepackage{graphicx}      
\begin{document}
\twocolumn[
\title{ALMA Observations of a High-density Core, MC27/L1521F in Taurus: Dynamical Gas Interaction at the Possible Site of a Multiple Star Formation}
\titlerunning{ALMA Observations of a High-density Core, MC27/L1521F in Taurus: Dynamical Gas Interaction at the Possible Site of a Multiple Star Formation}
\author{ Kazuki Tokuda }{tokuda@p.s.osakafu-u.ac.jp}
\address{Department of Physical Science, Graduate School of Science, Osaka Prefecture University,
1-1 Gakuen-cho, Naka-ku, Sakai, Osaka 599-8531, Japan}

\vskip 0.5cm 
]

\begin{abstract}
We present the results of ALMA observations of dust continuum emission and molecular rotational lines toward a dense core, MC27 (aka L1521F), which is considered to be very close to the first core phase. We revealed the spatial/velocity structures of the core are very complex and and suggest that the initial condition of star formation is highly dynamical.
\end{abstract}

\section{ALMA observations}
Our ALMA observations reveal complex structures at the center of MC27/L1521F (Tokuda {\it et al.} 2014). We find a few starless high-density cores, one of which (MMS-2) has a very high number density of $\sim$10$^7$ cm$^{-3}$. 
The high angular resolution observations at 0.87 mm with a beam size of $\sim$0.$''$74 $\times$ 0.$''$32 reveal that the MMS-2 has substructures in both dust and molecular emission (Fig. 1 lower panel).
A very compact bipolar outflow with a dynamical timescale of a few hundred years was found toward the protostar. 
The molecular line observations show several cores are associated with an arc-like structure whose length is $\sim$2000 AU, possibly formed due to the dynamical gas interaction (Matsumoto {\it et al.} 2015). 
The observed column density structure ranges in scale from $\sim$100 to $\sim$10,000 AU as revealed by
combining the 12 m array data with the 7 m array data as well as with the single dish MAMBO data (Fig. 1, see also Tokuda {\it et al.} 2016). 
The azimuthally averaged radial radial column density distribution of the inner part ($r$ $<$ 3000 AU) is $N_{\rm H_2}$ $\sim$$r^{-0.4}$, clearly flatter than that of the outer part, $\sim$$r^{-1.0}$. We detect the above-mentioned complex structures, such as the arc-like structures, inside the inner flatter region, which may reflect the dynamical status of the dense core (see also discussions in Tokuda {\it et al.} 2016). \\


\begin{figure}[h]
\vskip -0.5cm
\centering
$\begin{array}{cc}
\includegraphics[angle=0,height=16cm]{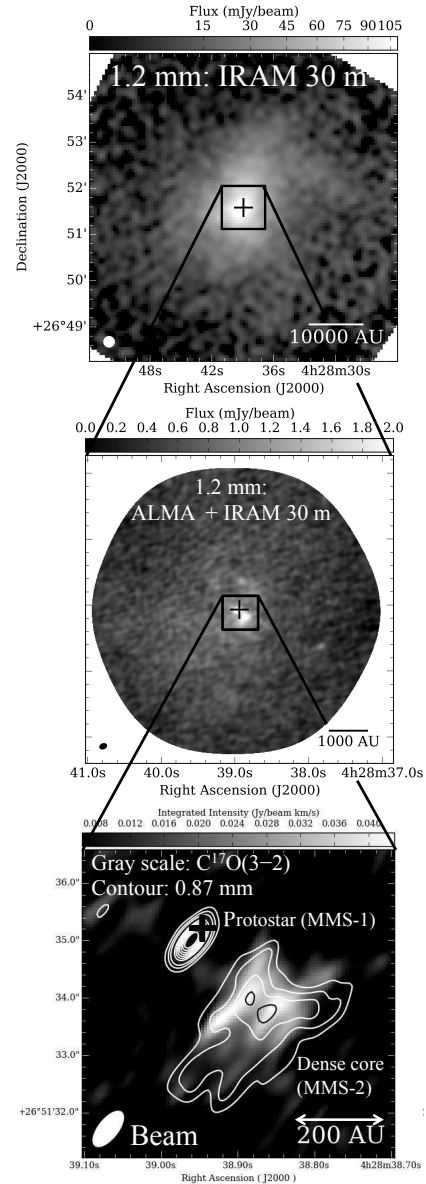} 
\end{array}$
\caption{Distribution of dust continuum and molecular line emission toward MC27/L1521F (Tokuda {\it et al.} 2016). Black cross in each panel represents the position of the protostar identified with $Spitzer$ observations.}
\label{fig:MC27}
\vskip -4.0cm
\end{figure}

\end{document}